\documentclass[useAMS,usenatbib]{mn2e}
\usepackage{graphicx}
\usepackage{natbib}
\usepackage{subfigure}
\usepackage{placeins}
\usepackage{amsmath}
\usepackage{rotating}
\usepackage{url}
\usepackage{amssymb}
\input{psfig.sty}

\newcommand{\aips}{{$\cal AIPS\/$}}

\def\l1.4{$L_{\rm 1.4GHz}$}
\def\s1.4{$S_{\rm 1.4GHz}$}

\def\kms{km\,s$^{-1}$ }

\def\jonezero{$J\!=\!1\!-\!0$}
\def\jthreetwo{$J\!=\!3\!-\!2$}
\def\jfourthree{$J\!=\!4\!-\!3$}

\def\jsixfive{$J\!=\!6\!-\!5$}
\def\jtennine{$J\!=\!10\!-\!9$}
\newcommand{\co}{$^{12}$CO}
\newcommand{\etal}{et~al.}
\newcommand{\msol}{\,\textrm{M}_{\odot}}                
\newcommand{\hst}{\textit{HST}}

\newcommand{\sfr}{M$_{\odot}$\,yr$^{-1}$}
\newcommand{\xco}{$\msol$\,({\sc k}\,km\,s$^{-1}$\,pc$^{2}$)$^{-1}$}
\def\gs{\mathrel{\raise0.35ex\hbox{$\scriptstyle >$}\kern-0.6em
\lower0.40ex\hbox{{$\scriptstyle \sim$}}}}
\def\ls{\mathrel{\raise0.35ex\hbox{$\scriptstyle <$}\kern-0.6em
\lower0.40ex\hbox{{$\scriptstyle \sim$}}}}
\def\m@th{\mathsurround=0pt }
\def\eqalign#1{\null\,\vcenter{\openup1\jot \m@th
 \ialign{\strut\hfil$\displaystyle{##}$&$\displaystyle{{}##}$\hfil
 \crcr#1\crcr}}\,}
\title[The ISM on $\sim 100$\,pc scales in a $z\sim 2.3$ galaxy]{Tracing cool molecular gas and star formation on $\sim 100$\,pc scales within a $z\sim 2.3$ galaxy}

\author[Thomson \etal]{A.\,P.\ Thomson,$^{\! 1}$\thanks{E-mail: alasdair.thomson@durham.ac.uk}
  R.\,J.\ Ivison,$^{\! 2,3}$
  Frazer N.\ Owen,$^{\! 4}$
  A.\,L.\,R\ Danielson,$^{\! 1}$ \and
  A.\,M.\ Swinbank,$^{\! 1}$
  Ian Smail,$^{\! 1}$
  \vspace*{1mm}\\
  $^{1}$Institute for Computational Cosmology, Durham University, South Road, Durham DH1 3LE\\
  $^{2}$Institute for Astronomy, University of Edinburgh, Blackford Hill, Edinburgh EH9 3HJ\\
  $^{3}$European Southern Observatory, Karl-Schwarzschild-Str. 2, 85748 Garching bei M\"{u}nchen, Germany\\
  $^{4}$National Radio Astronomy Observatory, P.O. Box O, Socorro, NM\,87801, USA\\
}
\date{Accepted 2015 January 6. Received 2014 December 24; in original form in 2014 October 30}
\pagerange{\pageref{firstpage}--\pageref{lastpage}} \pubyear{2014}

\begin{document}

\maketitle

\label{firstpage}
\begin{abstract}
We present new, high-angular resolution interferometric observations with the Karl G. Jansky Very Large Array of \co\ \jonezero\ line emission and $4$--$8$\,GHz continuum emission in the strongly lensed, $z=2.3$ submillimetre galaxy, SMM\,J21352-−0102. Using these data, we identify and probe the conditions in $\sim100$\,pc clumps within this galaxy, which we consider to be potential giant molecular cloud complexes, containing up to half of the total molecular gas in this system. In combination with far-infrared and submillimetre data, we investigate the far-infrared/radio correlation, measuring $q_{\rm IR}=2.39\pm 0.17$ across SMM\,J21352. We search for variations in the properties of the interstellar medium throughout the galaxy by measuring the spatially-resolved $q_{\rm IR}$ and radio spectral index, $\alpha_{\rm radio}$, finding ranges $q_{\rm IR}=[2.1, 2.6]$ and $\alpha_{\rm radio} = [-1.5,-0.7]$. We argue that these ranges in $\alpha_{\rm radio}$ and $q_{\rm IR}$ may reflect variations in the age of the ISM material. Using multi-$J$ \co\ data, we quantitatively test a recent theoretical model relating the star-formation rate surface density to the excitation of \co, finding good agreement between the model and the data. Lastly, we study the Schmidt-Kennicutt relation, both integrated across the system and within the individual clumps. We find small offsets between SMM\,J21352 and its clumps relative to other star-forming galaxy populations on the Schmidt-Kennicutt plot -- such offsets have previously been interpreted as evidence for a bi-modal star-formation law, but we argue that they can be equally well explained as arising due to a combination of observational uncertainties and systematic biases in the choice of model used to interpret the data.
\end{abstract}

\begin{keywords}
  galaxies: active --- galaxies: high-redshift --- galaxies: starburst ---
  submillimetre --- ISM: molecules --- galaxies: ISM
\end{keywords}

\section{Introduction}\label{sec:intro}

The resolution of the cosmic submillimetre background into discrete sources has revealed a population of high-redshift, ultra-luminous ($L_{\rm IR} \geq\ 10^{11-13}$L$_{\odot}$) galaxies \citep[e.g.][]{ivison98, walter12, riechers13}, the submillimetre galaxies (SMGs), whose bolometric luminosity is dominated by the re-radiance at infrared wavelengths of dust-obscured UV/optical photons produced by young stars and, in some cases, from accretion on to a central super-massive black hole, seen as an active galactic nucleus (AGN).

In spite of the intrinsically high luminosities of SMGs, many remain challenging targets for observational studies, due both to photon starvation in the 1\,mm atmospheric window commonly used to survey for them and the low angular resolution of these single-dish surveys. Indeed, with the exception of a handful of hyper-luminous objects (HyLIRGs), the brightest SMGs have 850-$\mu$m flux densities of only $\sim10$\,mJy \citep{ivison10a, karim13, barger14}. These effects conspire to suppress many examples of this important class of star-forming galaxy at or below the detection limit of confusion-noise limited single-dish surveys, hindering efforts to understand how and where they form their stars within their interstellar medium (ISM). To alleviate these issues, submillimetre surveys have been devised which exploit the panchromatic flux-boosting effects of gravitational lensing, either by individual line-of-sight foreground galaxies \citep[e.g.][]{rowan-robinson91, negrello10} or by deliberate or serendipitous exploitation of massive foreground clusters \citep{smail97, knudsen08, kneib10}.

Among the most striking star-forming galaxies detected in this manner is ``The Cosmic Eyelash'' (SMM\,J21352--0102; hereafter SMM\,J21352, at RA: $21\,35\,11.6$, Dec: $-01\,02\,52$, J\,2000), discovered by \citet{swinbank10a} following a survey of the $z=0.325$ MACS\,J2135-01 cluster with the Large Apex Bolometer Camera \citep[LABOCA, ][]{siringo09} on the 12-m Atacama Path-finder EXperiment (APEX) telescope. The high observed flux density of SMM\,J21352 -- $S_{870\, \mu{\rm m}} = (106\pm 7)$\,mJy -- is a result of the $\sim37.5\times$ magnification provided by the foreground cluster, and corresponds to a de-boosted flux density that is close to the confusion limit of submillimetre surveys ($S_{870 \mu {\rm m}}\sim 3$\,mJy). Its redshift, $z=2.3259$, derived through the first blind detection of \co\ \jonezero\ with the Zpectrometer instrument on the Robert C.\, Byrd Green Bank Telescope (GBT) is close to the median redshift ($z\sim 2.5$) of the SMG population \citep{chapman05, simpson14}, suggesting that SMM\,J21352 may be an intrinsically typical SMG, located serendipitously behind a strong cosmic lens. In addition to boosting its total flux above the confusion noise, the high magnification of SMM\,J21352 spatially extends the emission in the image plane, allowing individual star-forming regions within this high-redshift starburst galaxy to be identified and studied in detail.

Following the detection of SMM\,J21352, \citet{swinbank10a} conducted a follow-up programme of observations with the Submillimeter Array (SMA) at 870\,$\mu$m, where the $0.3''$ synthesized beam in the very extended (VEX) configuration resolved the emission in to eight distinct clumps over a $4''$ region, representing four unique dusty clumps of $\sim 100$\,pc size each (in the source plane), mirrored about the critical curve. 

The bulk molecular gas properties of SMM\,J21352 were first studied in \citet{danielson11}, in which a combination of single-dish \co\ \jonezero, mid-$J$ \co, C{\sc i} and HCN observations from the GBT, IRAM 30-m telescope and Plateau de Bure Interferometer (PdBI) were used to constrain the source-averaged excitation in SMM\,J21352. These authors have found evidence for a two-phase medium in SMM\,J21352 comprising a hot, dense, luminous component embedded within a massive ($>10^{10}$\,M$_{\odot}$), extended, cool, low-excitation reservoir. \citet{danielson11} use their spatially-integrated \co\ spectral line energy distribution (SLED) and spectrum to fit a kinematic model comprising four distinct clumps, with differing physical properties. Subsequently, \citet{swinbank11} studied the gas kinematics using moderate-resolution ($\sim0.8''$) observations from the Karl\,G.\,Jansky Very Large Array (VLA) of \co\ \jonezero\ emission, determining that they are well-described by a rotationally-supported disc model with an inclination-corrected rotation speed $v_{\rm rot}=320\pm 25$\,km\,s$^{-1}$, a rotational-to-dispersion velocity ratio $v/\sigma = 3.5\pm 0.2$ and a dynamical mass of $(6.0\pm 0.5)\times 10^{10}$\,M$_{\odot}$ within a 2.5\,kpc radius in the source plane. Later, \citet{danielson13} revisited SMM\,J21352 in a study of the optically-thin $^{13}{\rm CO}$ and C$^{18}$O lines, finding evidence of a C$^{18}$O over-abundance (relative to the Milky Way) in high-temperature ($>100$\,{\sc k}), dense ($n>10^3$\,cm$^{-3}$) regions, which may favour a somewhat top-heavy initial mass function.

Here, we present the latest high-resolution, high-sensitivity interferometric \co\ \jonezero\ map and spectral data cube of SMM\,J21352 obtained with the upgraded VLA, together with a new, deep VLA $C$-band continuum and spectral index map. We first use these observations to spatially-identify, and then study the excitation conditions within dense molecular gas clumps, providing an important independent check on previous results derived via kinematic decomposition of the source. In these clumps, we carry out the first observational test of recent theoretical models which relate the \co\ SLED directly to the star formation rate (SFR) density $\Sigma_{\rm SFR}$. Next, we use our new data to create a spatially-resolved Schmidt-Kennicutt (SK) plot, to search for evidence that either the clumps in SMM\,J21352 or the bulk ISM deviate strongly from the star-formation law ascribed to normal galaxies. We also study the spatially-resolved far-infrared/radio correlation (FIRRC), to search for evidence of possible age variations within the ISM.

In \S\,2, we present our observations and discuss the data reduction strategies for our new VLA data. In \S\,3 we present the latest constraints these data provide on the source-averaged properties of SMM\,J21352. \S\,4 utilizes the high angular resolution of our observations to measure the gas excitation conditions, radio properties and ratio of far-infrared/radio emission \textit{within} the ISM, and in \S\,5 we discuss the implications of these results for our understanding of star-formation at high redshift, and give our conclusions.

Throughout this paper, we use a cosmology with $H_{0}=71$\,km\,s$^{-1}$\,Mpc$^{-1}$, $\Omega_{\rm m} = 0.27$ and $\Omega_{\Lambda} = 0.73$, which gives an angular scale of $\sim 8.32/37.5=0.22$\,kpc\,arcsec$^{-1}$. All flux densities quoted are observed, and all luminosities -- unless otherwise stated -- are corrected for the $37.5\times$ amplification.

\section{Observations and Data Reduction}\label{sec:data-reduction}

\begin{figure*}
\centerline{\psfig{file=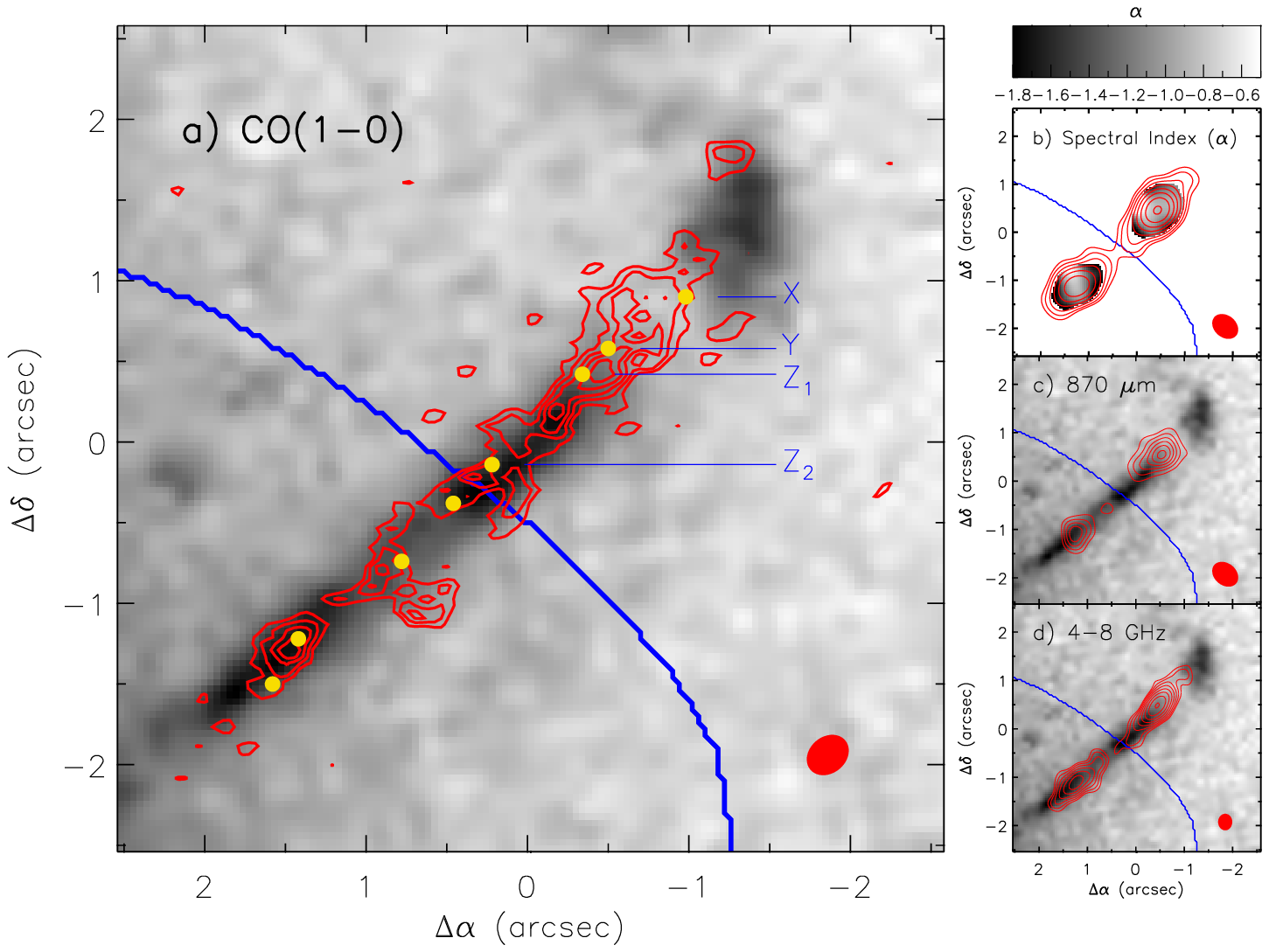,width=\textwidth}}
\vspace*{4mm}
\caption[\co\ \jonezero\ velocity-integrated images]{\textit{(a):} Velocity-integrated VLA \co\ \jonezero\ map (red contours), showing the position of the critical curve from the lensing model \citep[blue contour; ][]{swinbank10a}. The underlying greyscale is from the \hst/WFC3 {\sc $F140W$} observations of the foreground cluster, MACS\,J2135--0102 \citep{swinbank11}. Contours are spaced in intervals of $\sigma=14\,\mu$Jy, beginning at $3\sigma$. We mark with yellow dots the positions of the gas clumps identified in the \co\ \jonezero\ datacube via two independent methods (\S\,\ref{sect:mol-gas-clumps}), and identify them as spatially-resolved counterparts to the kinematic clumps $X$, $Y$, $Z_1$ and $Z_2$ of \citet{danielson11} \textit{(b): } Greyscale showing spatial variations in the $C$-band radio spectral index ($\alpha_{\rm radio}$). The red contours represent a spatially-smoothed version of the 4--8\,GHz data presented in the fourth panel. \textit{(c): } Updated SMA 870\,$\mu$m image, made using previously existing $uv$ data, but incorporating additional short spacings with respect to the VEX map presented in \citet{swinbank10a}. Contours are plotted in intervals of $\sigma= 1.4$\,mJy, beginning at $3\sigma$. \textit{(d): } High-resolution VLA $C$-band (4--8\,GHz) continuum image, with contours starting at $3\sigma$ (where $\sigma = 1.8\,\mu$Jy) and in steps of $\sqrt{2}\times\sigma$\ thereafter.  In all four maps, the beam used to create the contour image is shown at the bottom right.
}
\label{fig:eyelash-co}
\end{figure*}

\subsection{VLA \co\ \jonezero\ observations}
Observations of redshifted \co\ \jonezero\ emission in SMM\,J21352 were carried out using the VLA's new $Ka$-band receivers in 2--4\,hr blocks (totalling 25\,h of usable on-source data) in excellent weather conditions between 2010 September -- 2011 May. In \citet{swinbank11} we presented a datacube and map made from the hybrid DnC/CnB observations that were available at that time (VLA projects 10A-224 and 10B-195), in which the \co\ \jonezero\ emission was shown to be resolved on $\sim 0.8''$ scales -- here we incorporate the latest BnA-configuration (10C-141) observations, in order to maximize our angular resolution and study the ISM on hitherto inaccessible scales. During this Open Shared Risk Observing (OSRO) period, the Wide-band Interferometric Digital ARchitecture (WIDAR) correlator provided $64\times 2$\,MHz dual-polarization channels in each of two sub-band pairs, corresponding to a velocity resolution of $\sim 15$\,km\,s$^{-1}$\,channel$^{-1}$ before smoothing. 

Antenna pointings were checked hourly using the $C$-band receivers, and immediately prior to scans of the primary flux calibrator, 3C48. Amplitude and phase corrections as well as the bandpass were determined via short scans of the bright, nearby calibration source J2136+0041 in between each 5\,min integration on SMM\,J21352.

The data were flagged, calibrated and imaged using the standard \aips\ procedures outlined in \citet{ivison11}, though with a number of important changes: the data were downloaded in native {\sc smbdf} format and loaded into \aips\ using {\sc bdf2aips} rather than as {\sc uvfits} files using {\sc fitld}, in order to eliminate any server-side data compression artefacts. Known issues in the setting of delays at the VLA were compensated for in each track using {\sc fring}, and 1\,min of data towards the end of a good scan of 3C48.

For the DnC observations, the two sub-bands were tuned to 34.62 and 34.73\,GHz, such that at the GBT redshift \citep[$z=2.3259$; ][]{swinbank10a} the \co\ \jonezero\ line \citep[$\nu_{\rm rest} = 115.27120256$\,GHz; ][]{morton94} ought to have been centred in channel 50 of the lower sub-band (AC), with the upper sub-band (BD) tuned to overlap by 10-channels to mitigate the effects of noise in the end channels of each sub-band. The resulting 118 channels provide 2050\,km\,s$^{-1}$ of usable velocity coverage. In fact, spectra from these DnC-only data hinted at a slight redshift discrepancy with respect to the reference GBT redshift, and hence the array tunings for the subsequent CnB and BnA observations were shifted downward in frequency by 40\,MHz to accommodate the full width of the line. Thus, the usable spectral coverage common to all three array configurations is 98 channels, covering 1690\,km\,s$^{-1}$.

Calibrated $uv$ data for the three array configurations were concatenated together with {\sc dbcon}, with the {\sc reweight} parameter set to provide a 1:4 ratio in the sum of gridded weights between adjacent configurations, and so maximize the likelihood of the final data set producing a Gaussian synthesized beam when imaged.

We created maps and spectral data cubes of the \co\ \jonezero\ emission over a 3-arcmin field of view, which we cautiously cleaned by placing small, circular clean boxes around the brightest visible knots of emission after each major {\sc clean} cycle, then inspecting the residual image and iteratively adding more small clean boxes as appropriate until the network of small clean boxes overlapped across the full extent of the source and all emission lying significantly above the noise had been cleaned. Such conservative cleaning is necessary when creating interferometric maps of low surface brightness emission in order to prevent spurious emission (e.g. noise, or sidelobe emission) being carried through to the final {\sc clean} image, where it will lead morphological analysis astray.

We applied Gaussian $uv$ tapers from 50--1000\,k$\lambda$\ to the data in steps of 50\,k$\lambda$\ to search for any flux which may have been resolved out on long baselines, but found no evidence of significant missing flux. The un-tapered map (Figure.~\ref{fig:eyelash-co}) contains $2.07\pm0.01$\,mJy, $\sim95$\% of the single-dish GBT flux \citep{swinbank10a}, and hence we use this map without any additional tapering. The synthesized beam is $0.28\times 0.21$\,arcsec at position angle PA$=-50^{\circ}$.

\subsection{VLA C-band continuum observations}\label{sect:eyelash-radio}
SMM\,J21352 was observed in the extended A-configuration, in $C$-band (4--8\,GHz) at the VLA during 2011 July 6$^{\rm th}$--7$^{\rm th}$, in moderate weather conditions. These observations took place towards the end of the (Expanded) VLA hardware upgrade phase, by which time all 27 antennas had been fitted with functioning $C$-band receivers, and the WIDAR correlator was tuned to provide 16 sub-bands of $64\times 2$\,MHz channels each, yielding a total instantaneous bandwidth of 2\,GHz. By splitting the allocated time between observations at 4--6 and 6--8\,GHz, a moderately large spectral region of the radio continuum emission in SMM\,J21352 was studied.

From a total time allocation of 10\,h, 8\,h of usable on-source data were obtained, with flux calibration being set via scans of 3C48 at the end of each track, and antenna pointing, amplitude and phase tracking and bandpass correction being determined via frequent, short scans of the nearby calibrator J2136+0041.

These data were reduced and imaged in the Common Astronomy Software Applications ({\sc casa}\footnote{http://casa.nrao.edu}) package. Using the {\sc msmfs} {\sc clean} algorithm, we produce a high-resolution map with a synthesized beam $0.34''\times 0.29''$ at PA$=180^{\circ}$ and an rms sensitivity of 1.45\,$\mu$Jy\,beam$^{-1}$. The total flux density measured in this map is $S_{C}=0.258\pm 0.001$\,mJy. This map is shown in Figure\,\ref{fig:eyelash-co}.

\subsection{IRAM PdBI mid-$J$ \co\ observations}

We use observations from the six-element IRAM PdBI to study the mid-$J$ \co\ emission. These observations were originally presented in \citet{danielson11}. Using 4 and 2\,hr on-source exposures in D-configuration, taken in Directors Discretionary Time (DDT) during 2009 May, \co\ \jthreetwo\ and \jfourthree\ emission were observed, reaching signal-to-noise (S/N) $\sim 300$ across $900$\,km\,s$^{-1}$ in each velocity-integrated map. The total measured line fluxes are $13.20\pm0.10$\,Jy\,km\,s$^{-1}$ and $17.3\pm 1.2$\,Jy\,km\,s$^{-1}$ for \jthreetwo\ and \jfourthree\ respectively, and the corresponding synthesized beams are $2.11''\times1.59''$ at PA$=37^{\circ}$ and $1.08''\times 0.50''$ at PA$=23^{\circ}$. Further details of the data reduction and imaging are presented \citet{danielson11}, 

\co\ \jsixfive\ emission was observed in A configuration during 2010 January \citep[see][]{swinbank11}. The six hours of usable on-source integration time in this configuration yield a synthesized beam of $0.67''\times 0.43''$ at PA$=24.5^{\circ}$ and a velocity-integrated line flux of $21.5\pm1.1$\,Jy\,km\,s$^{-1}$.

\subsection{SMA observations}\label{sect:obs-sma}

A programme to map the rest-frame far-infrared continuum emission in SMM\,J21352 was carried out at the SMA in 2009 May -- September \citep[see ][in which a description of the data reduction is presented]{swinbank10a}. With the lower sub-band tuned to 870\,$\mu$m, a total of 5.5\,h of on-source data were obtained in the sub-compact (SC) configuration, with 12\,h in compact (C), 2\,h in extended (Ext) and 6\,h in the very extended (VEX) configuration. The upper sub-band was tuned to search for redshifted \co\ \jtennine\ emission -- this line was not detected, but a formal $3\sigma$\ upper-limit of $\leq 0.2$\,Jy\,km\,s$^{-1}$ was determined.

In order to search for signs of clumpy dust emission, \citet{swinbank10a} used only the six hours of VEX-configuration $uv$ data to produce a high-resolution $870\,\mu$m image. The resulting map -- with a $0.3''$ synthesized beam -- comprises eight, bright image-plane clumps (four source-plane clumps, each mirrored about the critical curve of the foreground lens), which account for $\sim 85\%$ of the total (i.e. single-dish LABOCA) $870\,\mu$m flux. In light of the updated information on the morphology of this system due to the new VLA $Ka$-band (covering \co\ \jonezero) and $C$-band (continuum) imaging, we opt to carry out a new imaging run from the existing calibrated SMA 870\,$\mu$m visibilities. We find that by incorporating previously unused visibilities from the Ext, C and SC array configurations -- though with a weighting scheme which still favours the longest baselines -- we suffer some degradation in resolution ($0.59''\times 0.45''$ at ${\rm PA}=50^\circ$, \textit{cf.} $0.29''\times 0.23''$ at ${\rm PA}=-15^\circ$ from VEX-only), but with the advantage that the improved sampling of the $uv$ plane yields a synthesized beam with less intrusive sidelobes. This new, more conservative SMA image (Figure\,1) has a sensitivity of $\sigma=1.4$\,mJy\,beam$^{-1}$, and the total flux density of $S_{870\,\mu{\rm m}}=91\pm5$\,mJy accounts for $\sim85$\% of the single-dish LABOCA flux reported in \citet{swinbank10a}.

\section{Source-averaged properties}

\subsection{\co\ \jonezero\ spectral line profile}

SMM\,J21352 is detected in \co\ \jonezero\ with a peak flux in the velocity-integrated image of $8\sigma$, where $\sigma=14$\,$\mu$Jy. We create a spectral data cube with 6\,MHz spectral resolution by binning the native VLA channels in groups of three in order to maximize signal-to-noise, and extract a line spectrum across the source using the \aips\ task {\sc blsum}. We see evidence of a broad line in the spectrum \citep[{\sc fwzi}$=920$\,km\,s$^{-1}$; in agreement with the existing mid-$J$ \co\ data from][]{danielson11}, with the velocity-integrated flux being consistent with that measured in the equivalent integrated image. This line profile places SMM\,J21352  at a flux-weighted redshift of $z=2.3268\pm0.0001$. The uncertainty in this redshift corresponds to velocity uncertainties of the order $\Delta$V$\sim 10$\,km\,s$^{-1}$. We produce an error spectrum using the {\sc rspec} task, in an off-source region of the cube. The typical noise in each 6\,MHz spectral channel is 51\,$\mu$Jy\,beam$^{-1}$.

\begin{figure}
\includegraphics[width=\columnwidth]{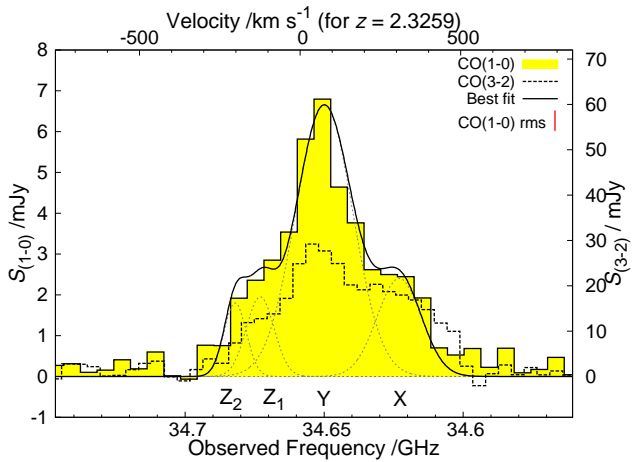}
\caption[SMM\,J21352 \co\ \jonezero\ and \jthreetwo\ spectra]{SMM\,J21352 \co\ \jonezero\ and \jthreetwo\ spectra. The latter is scaled vertically by a factor $9^{-1}$ so as to lie on the same $T_{\rm b}$ scale. The velocity scale is quoted with respect to the GBT \co\ \jonezero\ redshift, and a typical 50\,$\mu$Jy/channel error bar is shown in the top-right of the plot. The shapes of the two line profiles are in good agreement. The new VLA \jonezero\ line profile recovers $\sim95$\% of the flux of the GBT Zpectrometer \jonezero\ measurement (not shown) from \citet{swinbank10a}, but has a slightly different line profile due to the superior velocity resolution of the VLA's new $Ka$-band receivers relative to the coarser channels of the much-wider bandwidth Zpectrometer instrument on GBT. We fit this new \co\ \jonezero\ spectrum with emission from four kinematic components, which are well-modelled as a series of Gaussian components ($X$, $Y$ and $Z_1$ and $Z_2$: faint grey dotted lines).}
\label{fig:eyelash-spectrum}
\end{figure}

The peak \jonezero\ flux density seen in the source-averaged spectrum (Figure\,\ref{fig:eyelash-spectrum}) of $\sim7$\,mJy is slightly higher than the peak in the single-dish GBT spectrum presented in \citet{swinbank10a}, yet the velocity integrated flux densities agree within 5\%. We attribute the differences in line shapes between these line profiles to the $\sim 3\times$\ higher velocity resolution of the VLA data cube, and note that the new \jonezero\ line profile is similar to the shape of the \co\ \jthreetwo\ line profile presented in \citet[][,which we show in Figure\,\ref{fig:eyelash-spectrum} for comparison]{danielson11}, in which it is found that emission is better-fit by a superposition of Gaussian components with velocity offsets relative to the systematic zero velocity for SMM\,J21352 than they are by a single Gaussian component. Motivated by this analysis, we fit four Gaussians -- representing a possible interacting system -- over the {\sc fwzi}$=920$\kms\ width of the \co\ \jonezero\ line. The peaks and central velocities of these four components are given in Table\,\ref{tab:gaussian-fit}.

%
\begin{table}
  \begin{center}
  \caption[SMM\,J21352 kinematic parameters]{SMM\,J21352 kinematic parameters}
  \label{tab:gaussian-fit}
  \begin{tabular}{lcccc}
    \hline
    \multicolumn{1}{l}{Component} &
    \multicolumn{1}{l}{Peak} &
    \multicolumn{1}{c}{$v$} &\\
    \multicolumn{1}{l}{} &
    \multicolumn{1}{l}{(mJy)} &
    \multicolumn{1}{c}{(\kms)} &\\
    \hline
    $X$  & 2.4 & $+202\pm 30$\\
    $Y$  & 6.7 & $+120 \pm 30$\\
    $Z_1$& 2.0 & $-70 \pm 30$\\
    $Z_2$& 1.8 & $-310 \pm 30$\\         
    \hline
  \end{tabular}
  \end{center}
{\small

Notes: Best-fitting coefficients for the four-Gaussian fit to the observed \co\ \jonezero\ line. Velocities are quoted with respect to the reference redshift of $z=2.3259$, and uncertainties in the velocities are calculated by adding the statistical errors of the fit and the half-channel width in quadrature.}
\end{table}

\subsection{Line luminosity and total gas mass}\label{sect:eyelash-gas-mass}

Following \citet{solomon97}, the intrinsic (i.e. de-lensed) \co\ \jonezero\ line luminosity is given by

\begin{center}
\begin{equation}\label{eq:L_CO1}
L'_{\rm CO(1-0)} = \frac{3.25 \times 10^{7}}{\mu} \biggl[\frac{D_{\rm L}^{2}({\rm Mpc})}{1+z} \biggr] \biggl(\frac{\nu_{\rm rest}}{\rm GHz} \biggr) ^{-2}
\end{equation}
$\times \biggl[ \frac{\int_{\Delta V} S_{\nu}d\nu}{\rm Jy\,km\,s^{-1}} \biggr] {\rm K\,km\,s^{-1}\,pc^{2}}$
\end{center}

\noindent where $\mu$ is the average magnification of the source, $D_{\rm L}$ is the luminosity distance (Mpc), $\nu_{\rm rest}$ is the rest-frame central frequency of the line and $\int_{\Delta V} S_{\nu}d\nu$ is the velocity-integrated line intensity. We measure $L'_{\rm CO(1-0)}=(1.7\pm0.2)\times10^{10}$\,K\,km\,s$^{-1}$\,pc$^2$ -- in the limit that the molecular gas is optically thin, this implies a lower-limit on the (galaxy-averaged) \co-to-H$_2$ conversion factor\footnote{$\alpha_{\rm CO}$ includes a $\sim36$\% correction, based on primordial nucleosynthesis, for the contribution to the ISM of the second most abundant element, He. The molecular hydrogen mass is therefore $M_{{\rm H}_2}\sim M_{\rm gas}/1.36$.} $\alpha_{\rm CO}>0.54$, and hence $M_{\rm gas}>9.1\times10^9$\,M$_{\odot}$ \citep[see][]{ivison11, danielson11, danielson13}, while adopting the dynamical model of \citet{swinbank11} and assuming that the dynamical mass is entirely dominated by molecular gas\footnote{While such a scenario is physically implausible, we note the difficulty in accounting for the stellar mass in SMGs \citep[see][]{wardlow11, michalowski12}, and consider that our ``gas dominated dynamics'' toy model at least gives a firm upper limit on $M_{\rm gas}$.} gives corresponding upper-limits $\alpha_{\rm CO}<3.5$ and $M_{\rm gas}<6.5\times10^{10}$\,M$_{\odot}$.

\subsection{Radio spectral index}\label{sect:spec-index}

As a result of our strategy of splitting our VLA $C$-band observations between two different central frequencies, our observations cover a total bandwidth $\Delta\nu = 4$\,GHz, comprising 32 IFs of $64\times 2$\,MHz channels. This relatively large fractional bandwidth provides a long enough lever along which to measure the local radio spectral index. Setting the {\sc casa} {\sc clean} parameter {\sc nterms}$=2$ implements the multi-scale, multi-frequency synthesis ({\sc msmfs}) algorithm,  which uses two Taylor terms to model the sky emission, and simultaneously produces a cleaned continuum image, spectral index ($\alpha_{\rm radio}$) image, and spectral index error ($d\alpha_{\rm radio}$) image. The uncertainty, $d\alpha_{\rm radio}$ as a function of position in SMM\,J21352 is inversely proportional to the signal-to-noise in the continuum image -- we mask our spectral index image by excluding all pixels with $\left| d\alpha_{\rm radio} \right|>0.2$, which corresponds to a S/N cut of $\sim11\sigma$ in the continuum image.

The median spectral index of all pixels within this error threshold is $\alpha_{\rm radio}=-0.96\pm 0.13$, which is consistent with the integrated $Ka$-, $X$- and $C$-band flux densities measured in \citet{ivison10c}, which imply $\alpha_{\rm radio}=-0.22\pm 0.63$.

\begin{figure}
\centerline{\psfig{file=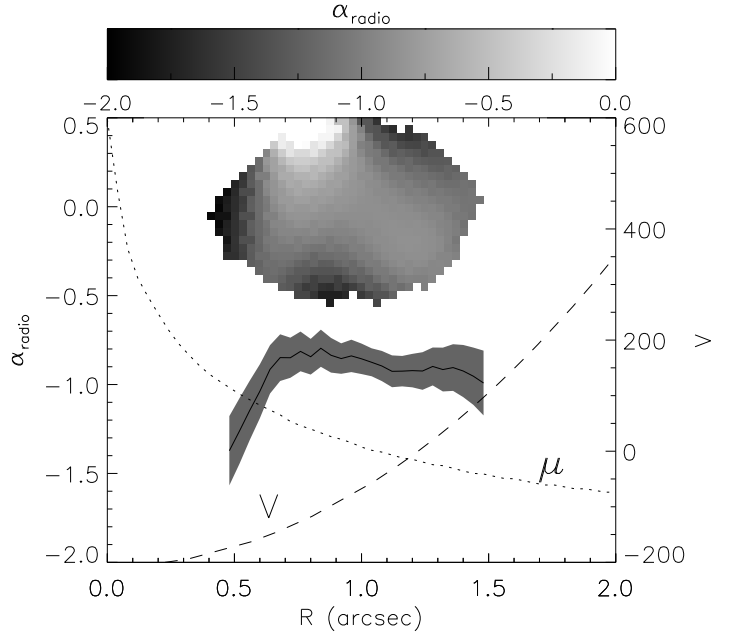,width=\columnwidth}}
\caption[Caption]{A plot showing variations in the radio spectral index, $\alpha_{\rm radio}$ as a function of distance ($R$) from the critical curve for the brighter of the two counter-images of SMM\,J21352. Shown at the top of the figure is a cut-out of the radio spectral index map for this counter-image, with greyscale running from black ($\alpha_{\rm radio}=-2.0$) to white ($\alpha_{\rm radio}=0.0$). The solid black line and shaded grey polygon represent the median and $\pm1\sigma$ scatter in $\alpha_{\rm radio}$ (respectively) as a function of distance from the critical curve. The dashed line represents the velocity field ($V$), derived from the \co\ \jonezero\ position-velocity diagram, while the dotted line represents (in arbitrary units) the variation in the strength of the magnifying effect with distance from the critical curve. We find that the bulk of the radio emission in SMM\,J21352 has $\alpha_{\rm radio}\sim-0.8$, consistent with a star formation origin, but that a highly-magnified pocket of emission $\sim 0.5''$ from the critical curve has considerably steeper spectral index, which may be the result of the on-going coalescence of two or more distinct components.}
\label{fig:eyelash-alpha-profile}
\end{figure}

The spectral index map is shown in Figure\,\ref{fig:eyelash-co}, with the positions of the \co\ \jonezero\ clumps marked for reference. We search for signs of variations in $\alpha_{\rm radio}$ with position along one dimension by collapsing the spectral index (and error) image(s) along an axis aligned with the position angle of SMM\,J21352 and measuring the median $\alpha_{\rm radio}$ and $d\alpha_{\rm radio}$ in each resolution step along this axis. We measure the spatially-reoslved gas velocity field by extracting a position-velocity (PV) diagram from the \co\ \jonezero\ datacube along the same slice, using the {\sc casa} task {\sc impv}. The $\alpha_{\rm radio}$ profile as a function of position from the critical curve for the brighter SMM\,J21352 counter-image is shown in Figure\,\ref{fig:eyelash-alpha-profile}. We see that radio emission located $0.7$--$1.5''$ from the critical curve is well described by $\alpha_{\rm radio}\sim -0.7$ (consistent with non-thermal synchrotron emission associated with star-formation), but that closer to the critical curve the radio spectrum noticeably steepens (to $\alpha_{\rm radio}\sim-1.5$). It is plausible that this region of steep-spectrum emission is the result of ongoing merger activity, where two (or more) star-forming progenitors spawn radio continuum bridges and tidal tails as the interaction proceeds. Such tidal tails are not associated directly with star-formation activity, but may produce ultra-steep spectrum ($\alpha_{\rm radio} < 1$) synchrotron emission as a result of relativistic electrons being shock-accelerated by the merger and confined within the interpenetrating magnetic fields of the progenitor galaxies \citep{lisenfeld10,murphy13}. We attempted to extract the radial $\alpha_{\rm radio}$ profile of the fainter counter-image as well, but found that the lower signal-to-noise of the data in this image rendered the results inconclusive except in the most strongly-lensed $\sim0.5''$ near the critical curve (where the results were consistent with, albeit with larger uncertainty than the results shown in Fig.\,\ref{fig:eyelash-alpha-profile}).

\section{Spatially resolved properties}\label{sect:eyelash-spatially-resolved-properties}
\subsection{Identification of molecular gas clumps}\label{sect:mol-gas-clumps}

By fitting a kinematic model to their source-integrated \co\ spectra, \citet[][2013]{danielson11} and \citet{swinbank11} discovered that much of the molecular gas in SMM\,J21352 resides in dense clumps in the galaxy. Initially, these were identified with clumpy structures seen in the SMA 870\,$\mu$m map of \citet{swinbank10a}, which appear to account for $\sim 80$\% of the total single dish flux. With our latest high-resolution \co\ \jonezero\ datacube, we check these positions using two independent methods.

First, we use the \aips\ task {\sc serch}, which employs the matched filter analysis technique outlined by \citet{uson91}, and attempts to fit Gaussian line profiles of user-specified channel width to each pixel in the datacube. Employing a 4$\sigma$ S/N cut and searching for lines between 50--950\,\kms\ (in steps of 100\,\kms), we identify eight bright clump features in the image plane (four to the north of the lens critical curve, and four mirror-images to the south), and use constraints provided by the lens model and by the velocity field to pair the eight image-plane clump features to four source-plane gas clumps. These gas clumps have slight positional offsets relative to the SMA clumps of \citet{swinbank10a}, but do have velocity centroids (relative to the systemic velocity of SMM\,J21352) that alow them to be identified with the kinematic clumps $X$, $Y$, $Z_1$, $Z_2$ of \citep{danielson11}.

The robustness of these clumps is verified by performing a blind search for clumps using the {\sc idl} clump-finding algorithm of \citet{williams94}\footnote{http://www.ifa.hawaii.edu/users/jpw/clumpfind.shtml}, which contours the data cube at $3\sigma$ (where $\sigma$ is the typical rms per spectral channel, $\sim 50$\,$\mu$Jy\,beam$^{-1}$) to identify peaks, and then attempts to follow the surrounding emission through adjacent beams until its extended structure attenuates below the noise threshold or blends with neighbouring clumps. We impose no prior image-plane or kinematic constraints on the algorithm, and the {\sc fwhm} of each clump line profile is computed on-the-fly, as the algorithm examines the cube. This algorithm reports 18 possible clumps in total, the brightest four of which agree positionally to within $\pm 0.5\times$ the synthesized beam with the four northern clumps identified via {\sc serch}. We identify four likely counterparts on the other side of the critical curve, within the same positional uncertainty. The positions of these clumps are shown in the main panel of Figure\,\ref{fig:eyelash-co}.

Though the latest, high-sensitivity (but lower resolution) SMA 870\,$\mu$m map fails to recover the clumpy structure seen in the previous VEX SMA map, it is noteworthy that we do observe clumpy structure in the \co\ \jonezero\ datacube whose kinematic properties are similar to those discussed in \citet{swinbank10a} and \citet[][whose analysis was guided by the VEX SMA image]{danielson11}, albeit with positional offsets between the \co\ and SMA-identified clumps of the order of $\sim1\times0.3''$ beam. Also noteworthy is the fact that these newly-identified clumps contain $\sim 10\%$ of the \co\ \jonezero\ line flux, with the majority of the cold gas residing in a more diffuse structure within which the clumps are embedded. 

If these \co\ gas clumps are marginally resolved by the VLA synthesized beam, then their physical sizes (once corrected for lensing) are $\sim100$\,pc, making them potential analogues of giant molecular clouds (GMCs), the sites of star formation in local galaxies. Do the physical conditions within the clumps justify this interpretation?

\begin{figure*}
\centerline{\psfig{file=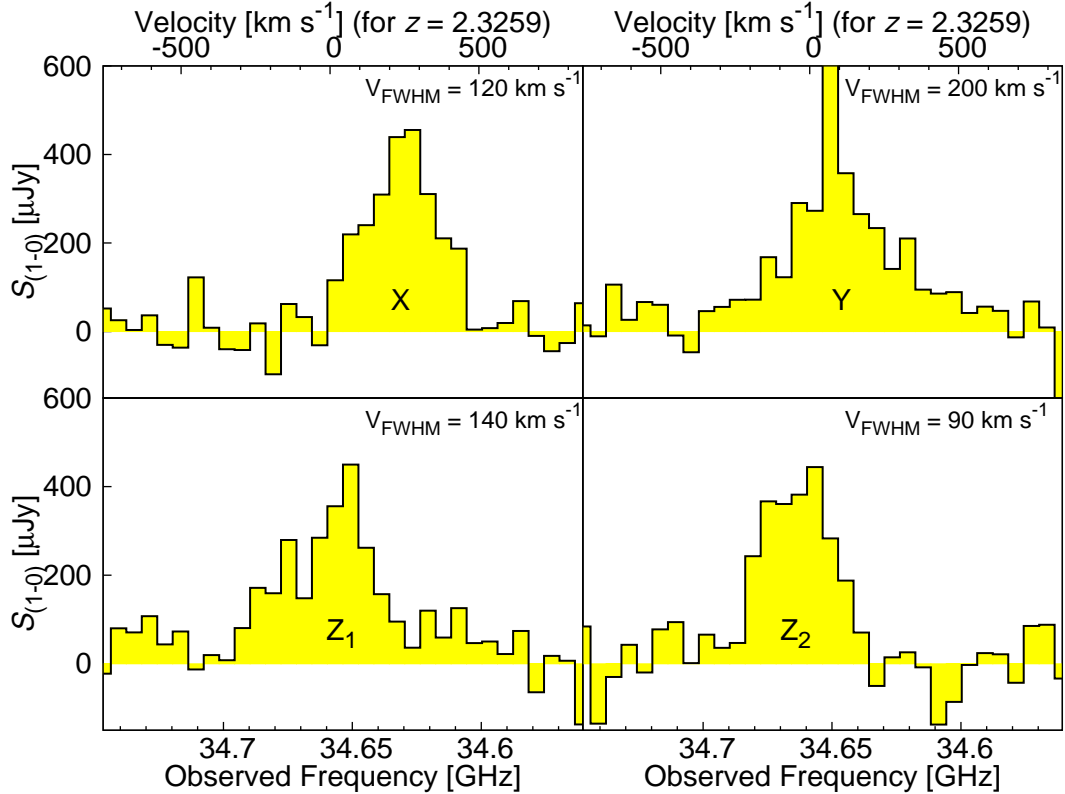,width=0.8\textwidth}}
\caption[SMM\,J21352 decomposed clump spectra]{\co\ \jonezero\ spectra for each of the four clumps identified via the methods presented in \S \ref{sect:eyelash-spatially-resolved-properties}. The velocity offsets seen between our spatially-resolved clumps allow them to be identified as likely counterparts to the kinematic clumps $X$, $Y$, $Z_1$ and $Z_2$, respectively, of \citet{danielson11}. The measured $V_{\rm FWHM}$ of each clump line profile is shown.}
\label{fig:eyelash-clump-spectra}
\end{figure*}

\subsection{Molecular gas excitation on $\sim100$\,pc scales}\label{sect:eyelash-radex}

We extract spectra from the \co\ \jonezero\ data cube at each of the clump positions. The resulting four clump spectra (Figure \ref{fig:eyelash-clump-spectra}) show velocity offsets with respect to the systematic velocity of SMM\,J21352. These offsets comprise one clump with a systematically positive velocity \citep[which we identify with the kinematically-resolved component $X$ of][]{danielson11}; the next clump, with a pseudo-Gaussian velocity profile with narrow {\sc fwhm} and broad {\sc fwzi} (which matches component $Y$ in the same paper); the remaining pair of clumps have systematically negative velocities (corresponding to clumps $Z_1$ and $Z_2$). 

The physical conditions of the gas within these clumps are probed by first determining the \co\ SLED of each clump, and then modelling the observed line ratios to constrain the temperature, density and molecular abundance in each clump.

We begin by placing the \co\ \jsixfive, \jfourthree, \jthreetwo, and \jonezero\ maps on a common astrometry/pixel scale using the \aips\ task {\sc ohgeo}, and deblending the flux in the mid-$J$ maps using a modified version of the algorithm presented in \citet{swinbank13}. Briefly, we seed model maps for each \co\ transition with eight delta functions at the locations of the clumps (and their mirror images) identified via our clump searching algorithms. The four clumps in the brighter mirror image north-west of the critical curve are constrained to have random flux densities between 0--150\% of the peak seen in the corresponding real integrated image, and the flux densities of the four fainter mirror images are constrained relatively via the lensing amplification map. Then, each model map is convolved with the appropriate synthesized beam, and the goodness-of-fit of that model map to the observed map is measured via the $\chi^2$ statistic, where the four degrees of freedom are those of the four independent clump fluxes:

\begin{center}
\begin{equation}
\chi^2 = \frac{N_{\rm pix}}{4} \times \frac{\Sigma ({\rm Image - Model})^2}{\Sigma ({\rm Model})^2}
\end{equation}
\end{center}

\noindent where $N_{\rm pix}$ is the number of pixels in each image (in each case, a $129\times 129$\,pix postage stamp with a pixel scale of 0.04-arcsec).  

After 500 fits (one \textit{major cycle}), we determine the set of clump fluxes for which $\chi^2$ reaches a minimum, re-centre on the best-fitting solutions and shrink the parameter space by 5\%, whereupon another set of 500 random clump fluxes are fit within the available parameter space. This process continues for 40 major cycles ($20,000$ fits in all), or until $\chi^2_{\rm min}$ has converged to within 1\% across three successive major cycles\footnote{In so doing, we implicitly assume that the emission from the $J_{\rm upper}>3$ states is confined to the dense clumps -- while this assumption is likely somewhat simplistic, we note that the high excitation conditions of the lines above $J=3$ ($n({\rm H}_2)\geq 10^4$\,cm$^{-3}$, $T_{k}\geq 50$\,{\sc k}) suggest that they preferentially trace star-forming regions, and render it unlikely that a substantial contribution to this emission arises from the diffuse ISM. To overcome this assumption will require future high-resolution observations of the mid-$J$ lines with ALMA.}.

Having thus determined the best-fitting \co\ SLEDs for each of the clumps, these are then used to constrain a large velocity gradient (LVG) model, computed using the {\sc radex} radiative transfer code \citep{vandertak07}. 

{\sc radex} computes the level populations, optical depths and intensities of each of the spectral lines of a chosen set of molecules under the assumptions that: (i) the medium is homogeneous; (ii) that collisional excitation of the species occurs only via interactions with H$_{2}$; and (iii) that the velocity widths of all molecular lines are equal for a given cloud. The collision rates used by {\sc radex} are functions of temperature, are held, along with other physical properties of each molecule, in a database of molecular files maintained by the Leiden Atomic and Molecular Database \citep[LAMBDA\footnote{http://home.strw.leidenuniv.nl/$\sim$moldata/};][]{schoier05}, and may be interpolated or extrapolated for $T_{\rm k}$ beyond the range contained within the datafile.

A suite of 640,000 model \co\ spectra were generated for each clump (with the line width altered appropriately in each case), for a grid of input parameters $T_{\rm k}$=15--90\,{\sc k} (in 16 steps of 5\,{\sc k}), n(H$_2$)=10$^{2-6}$\,cm$^{-3}$ (in 200 equidistant steps) and $N_{\rm CO} = 4\times10^{18-24}$\,cm$^{-2}$ (in 200 equidistant steps), values typical of molecular clouds seen in the local Universe \citep[e.g.][]{richardson85, ostriker10, shetty11}. In each model run, the background temperature was fixed at $T_{\rm CMB}=2.73\times(1+z)=9.08$\,{\sc k}, and the line-widths $\Delta V$ were set to the {\sc fwzi} of the individual clumps (Figure\,\ref{fig:eyelash-clump-spectra}).

\begin{figure*}
\centerline{\psfig{file=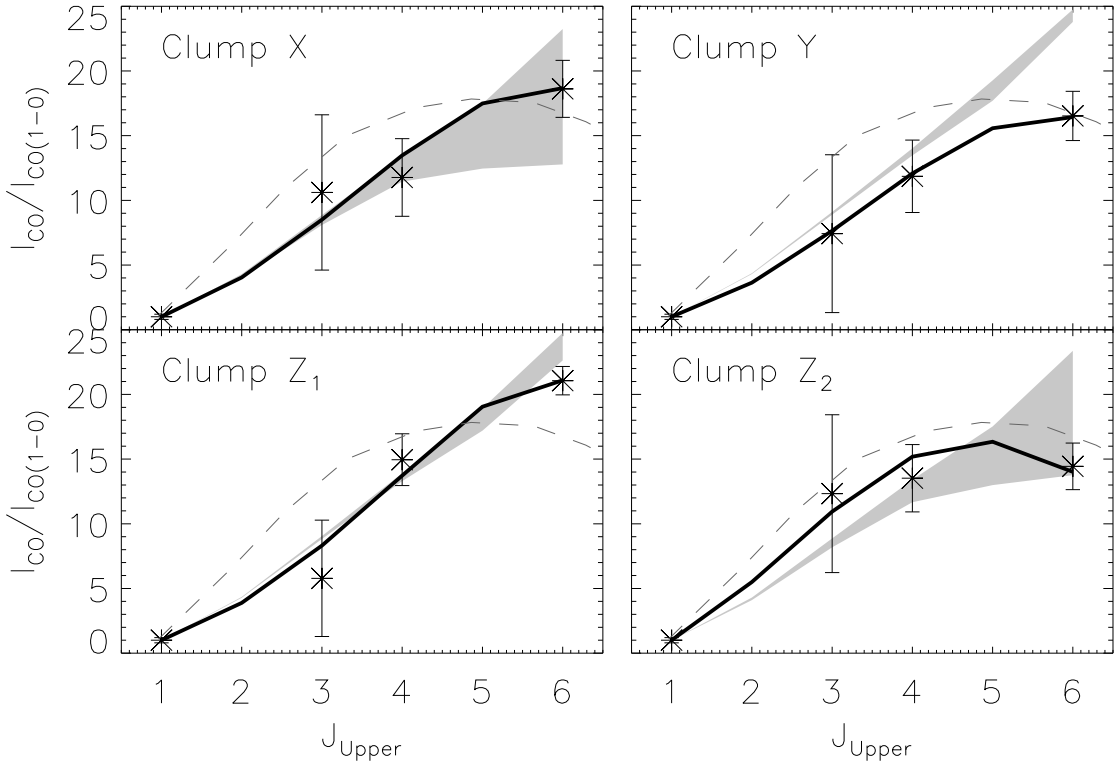,width=\textwidth}}
\vspace*{-10mm}
\caption[\co\ SLEDs of four resolved clumps in SMM\,J21352]{\co\ SLEDs of four marginally resolved clumps in SMM\,J21352, with the model SLEDs from {\sc radex} which best fit the data over-plotted as a solid black line. {\sc radex} models are generated across the parameter space $T_{\rm k}$=15--90\,{\sc k} (in 16 steps of 5\,{\sc k}), n(H$_2$)=10$^{2-6}$\,cm$^{-3}$ (in 200 equidistant steps) and $N_{\rm CO} = 4\times10^{18-24}$\,cm$^{-2}$ (in 200 equidistant steps). The key physical parameters of the best fitting (minimum-$\chi^2$) {\sc radex} models are given in Table\,\ref{tab:clump-properties}; we recover one clump with gas kinetic temperature $T_{\rm k} = 30$\,{\sc k} and three clumps with $T_{\rm k} = 70$\,{\sc k}, in good agreement with the distribution of dust temperatures measured by \citet{swinbank10a} and \citet{ivison10a}. The source-averaged \co\ SLED from \citet{danielson11} is shown as a dotted grey line for reference. Also shown (as shaded regions) are the $\pm2\sigma$ range of models from \citet[][see also \S\,\ref{sect:narayanan}]{narayanan14} which \textit{predict} the \co\ SLED, given estimates of $\Sigma_{\rm SFR}$ -- we see excellent agreement between the data and both models for clumps $X$, $Z_1$ and $Z_2$, however note that in clump $Y$ the \citet{narayanan14} model over-predicts the measured \co\ \jsixfive\ flux.}
\label{fig:eyelash-sled}
\end{figure*}

The measured \co\ SLEDs for each of the four clumps are shown with the corresponding best-fitting {\sc radex} models in Figure\,\ref{fig:eyelash-sled}. The shapes of the individual SLEDs are qualitatively consistent with the source-averaged SLED seen in \citet{danielson11}, which may be a consequence of the assumption that the bulk of the excited gas phase is contained within these clumps. The SLEDs peak for $J_{\rm upper} \geq 5$, in keeping with results from studies of local starburst galaxies \citep[e.g. M82 and NGC\,6240;][]{weiss05b,meijerink13}, AGN \citep[e.g. Mrk\,213; ][]{vanderwerf10}, and from the few high redshift SMGs thus-far to have yielded well-sampled SLEDs \citep[e.g. GN\,20 and SMM\,J14011;][respectively]{carilli10,sharon13}. The large error bars on the observed SLEDs are chiefly the result of degeneracies in the deblending process, which arise due to the poor angular resolution of the mid-$J$ PdBI maps.

From the best-fitting radiative transfer models for each clump, we compute $X_{\rm CO}={\rm N}_{{\rm H}{2}}/I_{\rm CO}$, the ratio of molecular hydrogen column density and \co\ \jonezero\ velocity-integrated line intensity, and then convert to the units of $\alpha_{\rm CO}$  via the conversion factor $\alpha_{\rm CO} (\msol\,({\rm K}\,{\rm km}\,{\rm s}^{-1}\,{\rm pc}^{2})^{-1}) = (2.17\times10^{-20})\times X_{\rm CO} ({\rm cm}^{-2}\,({\rm K}\,{\rm km}\,{\rm s}^{-1})^{-1})$, as given in \citet[][in which the $\sim36$\% contribution from helium is accounted for]{sandstrom13}.

The properties of the minimum-$\chi^2$\ fits to the data for the four clumps are summarized in Table \ref{tab:clump-properties}, and span the range $\log\bigl[{\rm n(H}_2)/{\rm cm}^{-3}\bigr]=3.6$--3.9, $\log\bigl[({\rm N_{\rm CO}/\Delta v)/{\rm cm}^{-2}km^{-1}s}\bigr]=15.7$--16.5. Three of the clumps have minimum-$\chi^2$\ temperatures  $T_{\rm k} \sim 70$\,{\sc k}, while the fourth appears to be significantly colder: $T_{\rm k} \sim 30$\,{\sc k}. If the temperatures of the gas and dust are coupled, then this range of gas temperatures is in agreement with the dual-component dust model \citep{ivison10a} which fits the far-IR SED of SMM\,J21352. We briefly note that our measured H$_2$ densities (n(H$_2$)) are concordant with those measured by \citet{danielson13} in their analysis of optically-thin $^{13}$CO and C$^{18}$O emission, while our temperatures are considerably lower than their range of best-fitting temperatures ($T_{k}=140$--$200$\,{\sc k}). However the uncertainties on these measurements are large, and at least for clumps $X$, $Z_1$ and $Z_2$, our measured temperatures are within the $\Delta\chi^2 = 1\sigma$ range quoted by \citet{danielson13}. Only for component $Y$ is our best-fitting temperature formally outside their range of allowed temperatures ($T_{k}=110$-$200$\,{\sc k}). The most likely explanation for this apparent disagreement is that although the spatially-identified \co\ \jonezero\ gas clumps have similar velocity offsets (relative to the systemic velocity of SMM\,J21352) to those clumps studied in \citet{danielson13}, the differing methods by which the two sets of clumps have been identified mean that we may not be looking at \textit{exactly} the same material in both cases. The low excitation requirements of the \co\ \jonezero\ line make it ubiquitous throughout the cool ISM -- thus regions with high \jonezero\ surface brightness correspond to regions of dense, cold gas -- while both C$^{18}$O and $^{13}$CO preferentially trace regions of recent, intense star-formation \citep[e.g.][]{meier04}, in which we may expect to see an elevated gas temperature.

Since {\sc radex} computes molecular line intensities for a given average \textit{volume} density, n(H$_2$), we must integrate the mean density along a line of sight, $\int ds$, through each cloud in order to calculate the column density: N$_{{\rm H}_{2}}$ = $\int$n(H$_2$)$ds$.

%
\begin{table*}
  \centering
  \caption[SMM\,J21352 clump-scale gas properties]{SMM\,J21352 clump properties}
  \label{tab:clump-properties}
  \begin{tabular}{lcccccccc}
    \hline
    \multicolumn{1}{l}{Clump} &
    \multicolumn{1}{c}{$T_{k}$} &
    \multicolumn{1}{c}{$\log\bigl[{\rm n(H}_2)$} &
    \multicolumn{1}{c}{$\alpha_{\rm CO}$} & 
    \multicolumn{1}{c}{$M_{\rm gas}$} & 
    \multicolumn{1}{c}{$\Sigma_{\rm SFR}$} &
    \multicolumn{1}{c}{$\alpha_{\rm radio}$} &
    \multicolumn{1}{c}{$q_{\rm IR}$} &\\
    \multicolumn{1}{l}{} &
    \multicolumn{1}{c}{({\sc k})} &
    \multicolumn{1}{c}{$/{\rm cm}^{-3}\bigr]$} &
    \multicolumn{1}{c}{({\sc k}\,km\,s$^{-1}$\,pc$^{2}$)} &
    \multicolumn{1}{c}{($\times 10^{9}$\,M$_{\odot}$)} & 
    \multicolumn{1}{c}{\sfr\,kpc$^{-2}$} &
    \multicolumn{1}{c}{} &
    \multicolumn{1}{c}{} &\\
    \hline
    $X$&70&$3.9\pm0.4$  &$<7.8$ &$<5.4$&$490\pm 220$ &     --       &$2.5\pm 0.2$\\
    $Y$&70&$3.6\pm0.4$  &$<0.6$ &$<6.1$&$1650\pm 310$&$-0.88\pm0.07$&$2.3\pm 0.1$\\
    $Z_1$&30&$3.7\pm0.6$&$<9.4$ &$<7.8$&$1360\pm 280$&$-0.82\pm0.07$&$2.2\pm 0.1$\\
    $Z_2$&70&$3.9\pm0.1$&$<16.1$&$<10.6$&$510\pm 220$ &     --       &$2.6\pm 0.2$\\
    \hline
  \end{tabular}
\end{table*}

This conversion from volume density to column density of H$_2$ implies that the best $\alpha_{\rm CO}$ is strictly an upper limit; the only estimate available for the sizes of the clumps ($ds$) is the constraint that they be no larger than the area covered by VLA synthesized beam ($\sim 100$\,pc). If the clumps are in fact significantly more compact than this, then $\alpha_{\rm CO}$ will scale downward correspondingly. The resulting limits on $\alpha_{\rm CO}$ are reported in Table\,2.

The measured upper-limits in $\alpha_{\rm CO}$ for the clumps are consistent with the range of Ultra-Luminous Infrared Galaxy (ULIRG)-like to Milky Way GMC-like values, and, in conjunction with estimates on $L'_{\rm CO}$ allow upper-limits to the gas masses to be determined (Table\,\ref{tab:clump-properties}). Depending on whether the total molecular gas mass of SMM\,J21352 is closer to the firm lower or upper limits determined in \S \ref{sect:eyelash-gas-mass}, these clumps may between them contain 10--60\% of the total molecular gas budget of the system.

\subsubsection{Comparison with recent theoretical work}\label{sect:narayanan}

Recently, \citet{narayanan14} presented a theoretical framework for the excitation of \co\ in star-forming galaxies which draws together numerical simulations of disc and merging galaxies with molecular line radiative transfer calculations. Within this framework, it is predicted that although in general the shape of the \co\ SLED is determined by several physical properties which are observationally challenging to determine (such as the gas density, temperature and optical depth in the molecular ISM), the fact that each of these properties correlates in theoretical models in some predictable way with the SFR surface density, $\Sigma_{\rm SFR}$, allows the \co\ SLED to be predicted given a measure of $\Sigma_{\rm SFR}$.

With our latest SMA 870\,$\mu$m map, we estimate $\Sigma_{\rm SFR}$ for each of the four clumps. We begin by measuring the flux density ($S_{\rm 870\,\mu{\rm m}}$) of each clump, and fitting a scaled version of the SMM\,J21352 far-IR template \citep{ivison10a}, which we integrate between $8$--$1000\,\mu$m to get $L_{\rm IR}$. We adopt the $L_{\rm IR}$ to SFR conversion factor of \citet{kennicutt98}, and estimate clump sizes based on the angular size of the SMA beam (corrected for amplification). 

We use an {\sc idl} script to generate for each clump $10^5$ Normally-distributed values of $\Sigma_{\rm SFR,model}$, whose central value and standard deviation are given by the measured $\Sigma_{\rm SFR}$ for that clump, and its uncertainty. We then call the \citet{narayanan14} {\sc idl} code for each $\Sigma_{\rm SFR,model}$ in the distribution, giving a library of 100,000 \co\ SLEDs for each clump. We show the range of model SLEDs whose input $\Sigma_{\rm SFR,model}$ is within $\pm2\sigma$ of its measured value for each clump as a greyscale in Figure\,\ref{fig:eyelash-sled}, along with the observed SLEDs from our deblending work, and the best-fitting model SLEDs from our {\sc radex} radiative transfer calculations. We see excellent agreement between both sets of models and the data for clumps $X$, $Z_1$ and $Z_2$, however the \citet{narayanan14} SLED for clump $Y$ diverges slightly at $J>4$. Hence, to the extent that our low-resolution mid-$J$ \co\ data can constrain the clump SLEDs, we find that the \citet{narayanan14} models adequately predict the \co\ SLEDs of star-forming regions for a given $\Sigma_{\rm SFR}$ around 75\% of the time.

\subsection{The FIRRC and radio spectral index}\label{sect:firrc}

It has long been established that in local star-forming galaxies, there exists a correlation between the relative strengths of far-infrared and radio emission \citep[e.g.][]{helou85, condon91} on scales from GMC complexes up to entire galaxies. Furthermore, this correlation is observed to hold ubiquitously over several orders of magnitude in luminosity, gas surface density and photon energy density.

This FIRRC is most readily understood within the framework of calorimetry \citep[e.g.][]{voelk89, lacki10}, in which UV and optical photons produced by young, massive stars are captured by dust and re-radiated at infrared wavelengths at a rate that is closely matched by that at which the same massive stars produce synchrotron emission via supernova shocks when they die. Provided the duration of the star formation event is greater than the $\sim 10$\,Myr time-scale on which the first supernovae are produced, then the two processes can reach equilibrium, giving rise to the observed correlation.

The FIRRC is commonly expressed \citep[as in][]{ivison10c} via the single logarithmic parameter $q_{\rm IR}$, given as

\begin{center}
\begin{equation}
q_{\rm IR} = \log_{10}\biggl[{
\frac{S_{\rm IR}}{3.75\times 10^{12}{\rm W\,m}^{-2}}
\times \frac{{\rm W\,m}^{-2}{\rm \,Hz}^{-1}}{S_{\rm 1.4\, GHz}}
}\biggr]
\end{equation}
\end{center}

\noindent where $S_{\rm IR}$ is the far-infrared (8--1000\,$\mu$m) flux, and $S_{1.4\,{\rm GHz}}$ is the monochromatic 1.4\,GHz radio flux, $k$-corrected to the rest-frame.

We investigate this interplay between thermal dust emission and non-thermal synchrotron emission in SMM\,J21352 by making a tapered $C$-band radio map whose resolution is close to that of the SMA map, and then convolving this image with a Gaussian kernel to exactly match the SMA resolution. Next, we regrid both images, binning by a factor $3\times$\ in both dimensions, and $k$-correct the total measured $C$-band radio flux density in the regridded map (using the average spectral index, $\alpha_{\rm radio}=-0.96$) to $\nu_{\rm rest}=1.4$\,GHz. Using $L_{\rm IR}=(2.3\pm0.2)\times10^{12}$\,L$_{\odot}$ as in \S\,\ref{sect:narayanan}, we measure the galaxy-averaged FIRRC parameter $q_{\rm IR}=2.39\pm0.17$.

Next, we capitalize on the high spatial resolution of our new dataset by measuring $q_{\rm IR}$ on a (rebinned) pixel-by-pixel basis throughout the molecular gas reservoir. We use the rebinned radio and spectral index maps to create a $k$-corrected rest-frame 1.4\,GHz flux density map, and create a map of $S_{\rm IR}$ by scaling the observed infrared SED for SMM\,J21352 \citep{swinbank10a, ivison10a} to fit the measured 870\,$\mu$m flux density at each position. We integrate at each pixel between $8$--$1000$\,$\mu$m to obtain a spatially-resolved $L_{\rm IR}$ map. In so doing, we explicitly assume that the far-IR SED of each resolution element is representable by the same mixture of 30\,{\sc k} and 70\,{\sc k} dust as explains the SMM\,J21352 SED. We restrict our analysis to pixels which sit within the $3\sigma$ contour of \co\ \jonezero\ emission, and for which the uncertainty in the spectral index $d\alpha_{\rm radio}\leq 0.2$. The median value of $q_{\rm IR}=2.33\pm0.02$ is consistent with that derived for SMM\,J21352 using the total flux densities.

\begin{figure*}
\includegraphics[width=\textwidth]{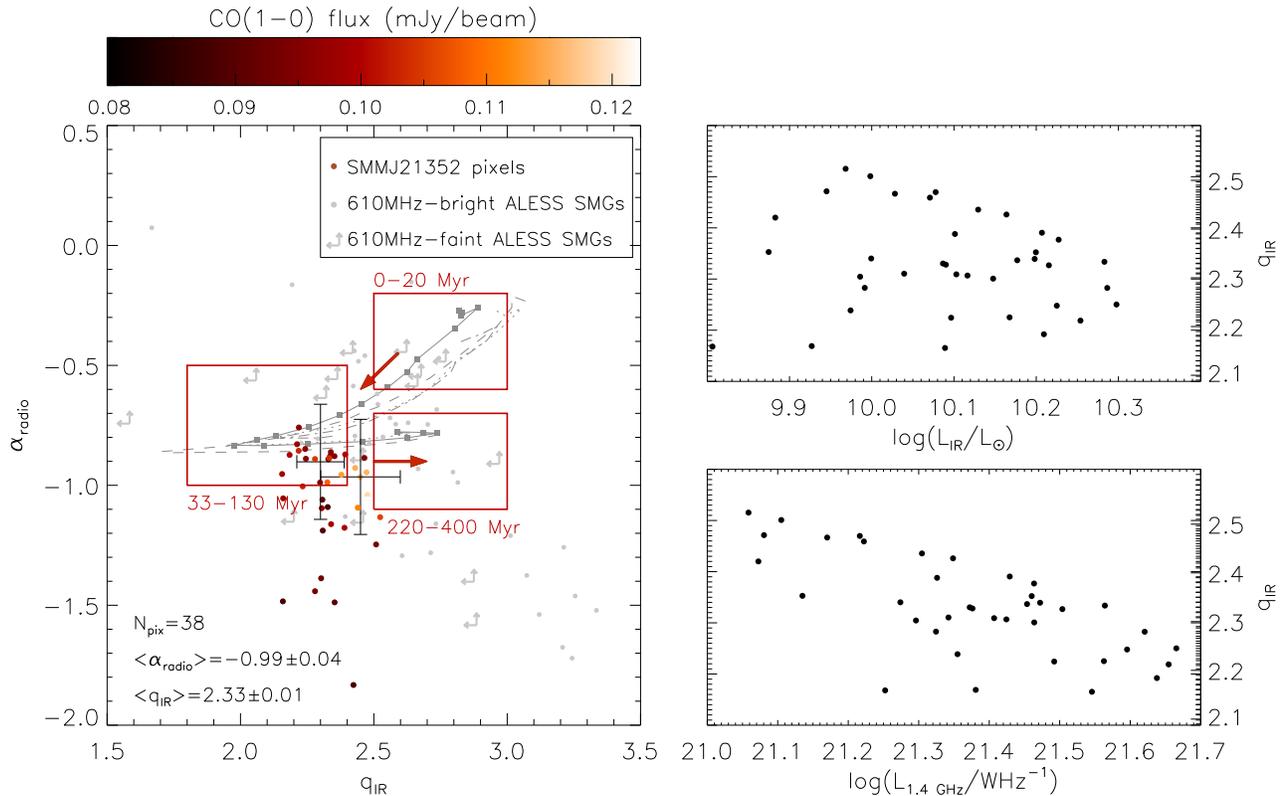}
\caption[Radio spectral index v $q_{\rm IR}$]{\textit{Left:} Plot of spatially resolved (i.e. pixel-by-pixel) variations in the radio spectral index, $\alpha_{\rm radio}$, versus FIRRC parameter, $q_{\rm IR}$ (with the error bars shown for clumps $Y$ and $Z_1$ being representative for the majority of pixels). Data points are colour-coded by the \co\ \jonezero\ flux density from the smoothed VLA map. We see a tendency for dense gas (high \jonezero\ flux; lighter-coloured points) to be co-spatial with regions of higher $q_{\rm IR}$ (i.e. enhanced infrared emission relative to radio), which is expected if the densest gas is co-spatial with the youngest star-forming regions which have not yet produced significant non-thermal radio emission from supernovae. For comparison, we plot (in grey) the results of a similar analysis of individual ALMA-observed SMGs from the on-going ALESS programme \citep[data taken from][]{thomson14}. Also plotted are model tracks for evolving starburst galaxies \citep[][see \S\,\ref{sect:bressan} for details]{bressan02}, with red boxes denoting the predicted ages of the galaxies/material they contain. \textit{Top-Right:} Spatially-resolved $q_{\rm IR}$ versus $\log({\rm L_{\rm IR}})$ and \textit{Bottom-Right:} $q_{\rm IR}$ versus $\log({\rm L_{\rm radio}})$, from which we see that that spatial variations in $q_{\rm IR}$ across the surface of SMM\,J21352 are driven predominantly by the radio morphology of the source.}
\label{fig:alphavqir}
\end{figure*}

We present the pixel-by-pixel distribution of $\alpha_{\rm radio}$\ and $q_{\rm IR}$ within the molecular gas reservoir in Figure\,\ref{fig:alphavqir}. Data are shown as coloured dots (colour-coded by \co\ \jonezero\ flux density), with grey symbols representing measurements and limits on $\alpha_{\rm radio}$ and $q_{\rm IR}$ in individual SMGs from our programme of ALMA observations in the Extended \textit{Chandra} Deep Field South \citep[ALESS: ][]{hodge13, thomson14}. Also plotted in Figure\,\ref{fig:alphavqir} are model tracks from \citet{bressan02}, along which starburst galaxies are expected to evolve from young to old over a timescale of $\sim 400$\,Myr. The pixels at the positions of clumps $Y$ and $Z_1$ are highlighted with error bars (which are representative of the error bars for all pixels). Clumps $X$ and $Z_2$ lie in the masked region of the spectral index image, for which $d\alpha_{\rm radio}>0.2$, and thus we are unable to report their $\alpha_{\rm radio}$. We see that our rebinned pixels in SMM\,J21352 span a narrow range in $q_{\rm IR}$ (possibly a direct consequence of the assumption of a common dust SED for all pixels), but that they exhibit considerable scatter in $\alpha_{\rm radio}$, intersecting the model tracks for evolving starburst galaxies of \citet{bressan02}.

\subsection{The Schmidt-Kennicutt Law}

A critical ingredient in any attempt to understand galaxy formation and evolution is a proper characterization of the processes by which diffuse baryonic matter comes to be locked up in stars. Unfortunately, the spatial scales on which star formation actually occurs remain beyond observational reach for all but the most nearby galaxies, and thus efforts to characterize these processes in distant galaxies are limited to the application of simplified recipes, or star-formation laws. 

It has long been claimed \citep[e.g.][]{greve05} that SMGs follow a Schmidt-type star-formation law,  $\Sigma_{\rm SFR}\propto\Sigma_{\rm gas}^{N}$ (where $\Sigma_{\rm SFR}$ and $\Sigma_{\rm gas}$ are the SFR and molecular gas surface densities, respectively, and $N$ is the power-law index), albeit possibly with an enhanced SFR per unit molecular gas relative to the population of normal star-forming galaxies; this has been interpreted as evidence of a dual star-formation law at work \citep{daddi08, daddi10b, dannerbauer09}. Under this paradigm, photometrically-selected $BzK$ galaxies \citep{daddi04} typify the main sequence of ongoing star-formation activity, fuelled by cold mode accretion from the surrounding intergalactic medium, while SMGs form stars in shorter-lived, hyper-efficient spurts due to major merger activity \citep[e.g.][]{genzel10}. However, many of the studies which have been claimed to corroborate this model have relied on observations of the mid-$J$ \co\ lines, which, while brighter (and hence easier to observe) offer a view of the molecular gas reservoir that is biased by their excitation requirements (i.e. denser, warmer gas, which may not be representative of the \textit{bulk} of gas reservoir). Such observations are reliant on an assumed gas excitation model, from which $\alpha_{\rm CO}$ (and ultimately $M_{\rm gas}$) is inferred \citep[see][for further discussion of this point]{ivison11}.

The high angular resolution of our new VLA maps for SMM\,J21352 allow us to evaluate the spatially-resolved Schmidt-Kennicutt relation in a high-redshift star-forming galaxy in \co\ \jonezero, without the need to assume a gas excitation model. Tapering the \co\ \jonezero\ $uv$ data and convolving the tapered image with a Gaussian kernel, we create a map of molecular gas emission with the same resolution as the updated 870\,$\mu$m map, and measure $L'_{\rm CO(1-0)}$ in each of the dense gas clumps.

We use estimates of the SFR of each clump from \S\,\ref{sect:firrc}, and convert $L'_{\rm CO(1-0)}$ to $M_{\rm gas}$  via a suitable choice of $\alpha_{\rm CO}$. The lowest $\alpha_{\rm CO}$ produced by our radiative transfer modelling of the individual clumps (Table\,\ref{tab:clump-properties}) is $\alpha_{\rm CO}=0.6$\xco, in clump $Y$. For $X$, $Z_1$ and $Z_2$, the $\alpha_{\rm CO}$ upper limits are consistent with the galaxy-averaged $\alpha_{\rm CO} \leq 3.5$ (\S\,\ref{sect:eyelash-gas-mass}), and so we adopt this limit for these three clumps. It will not be possible to significantly improve upon this assumption until even higher resolution \co\ observations allow us to better estimate the sizes of the clumps, and thus better constrain the conversion from H$_2$ column density to number density. 

\begin{figure}
\centering
\includegraphics[width=\columnwidth]{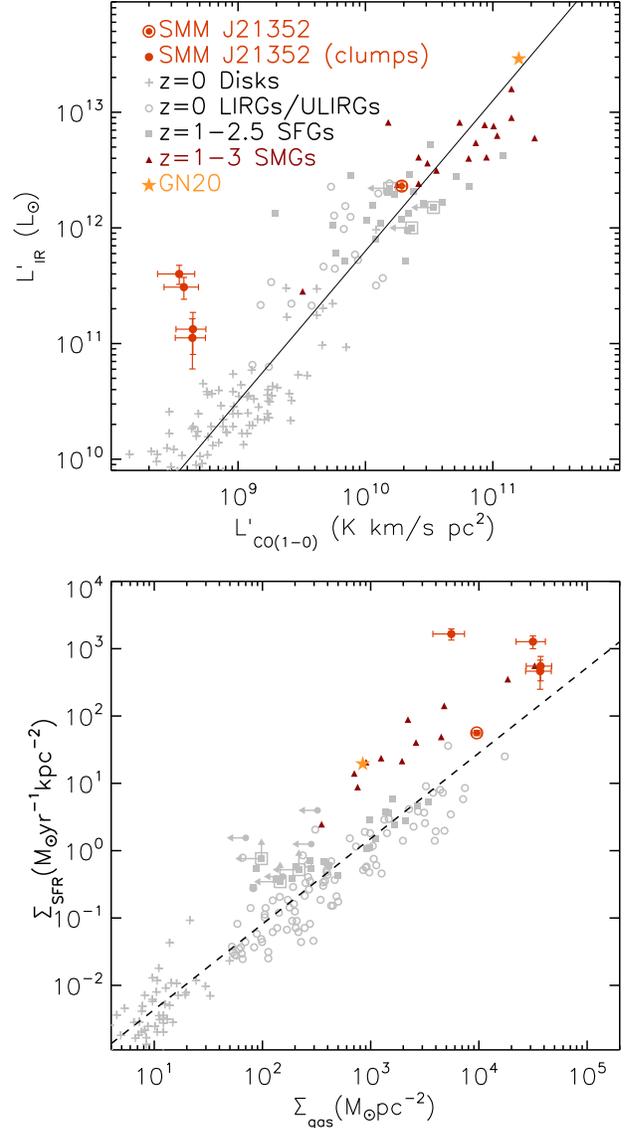}
\vspace*{-12mm}
\caption[The Schmidt-Kennicutt plot at high-$z$]{The Schmidt-Kennicutt plot, relating star formation and the availability of molecular gas. Local discs, LIRGs, ULIRGs and high-redshift star-forming ($BzK$) galaxies are taken from \citet{genzel10}. SMGs with recent \co\ \jonezero\ measurements, including the GN\,20 object of \citet{carilli10} are shown. \textit{Top: } The observed S-K relation. After correction for lensing, SMM\,J21352 is seen to be an intrinsically typical SMG, while the analysis of the clumps extends the plot at high redshift to luminosities hitherto well below the confusion limit. \textit{Bottom: } The surface density S-K relation, in which we convert the observed \co\ \jonezero\ luminosities in to gas masses using a standard ULIRG-like $\alpha_{\rm CO}=0.8$ for the whole galaxy (to facilitate comparison with the literature) and $\alpha_{\rm CO}=0.67$ for the clumps, based on the radiative transfer calculations presented in \S\,\ref{sect:eyelash-radex}. On both plots, the error bars shown are calculated from the signal-to-noise of the SMA 870\,$\mu$m and \co\ \jonezero\ images, on the assumption that the clump sizes are well-known. If the clumps are in fact much  \textit{smaller} than the VLA beam, then both $\Sigma_{\rm SFR}$ and $\Sigma_{\rm gas}$ are raised proportionally, but with an additional raising of $\Sigma_{\rm gas}$ due the lower $\alpha_{\rm CO}$ that this increase in surface density would imply. The dense clumps in SMM\,J21352 populate the high-density tail of the so-called ``merger sequence'', however systematic uncertainties in $\alpha_{\rm CO}$ in both $BzK$ galaxies and SMGs make the distinction between these two sequences unclear.}
\label{fig:sk-eyelash}
\end{figure}

Acknowledging the limitations imposed by these various assumptions, the measured $\Sigma_{\rm gas}$ and $\Sigma_{\rm SFR}$ of the clumps are consistent with an extrapolation of the Schmidt--Kennicutt Law (Figure\,\ref{fig:sk-eyelash}), making them among the densest star-forming environments seen at high-redshift. SMM\,J21352, too, unsurprisingly appears to obey the Schmidt--Kennicutt Law, albeit at lower surface density. The positioning of the clumps on the $\Sigma_{\rm SFR}/\Sigma_{\rm gas}$ plot sees them slightly offset from the star-formation law in ``normal'' galaxies, but is subject to the uncertainties in the values of $\alpha_{\rm CO}$ used to derive them.

\section{Discussion and conclusions}

\subsection{New constraints on the  spatially-resolved \co\ SLED}
We have presented new, high resolution VLA observations of \co\ \jonezero\ and C-band 4--8\,GHz continuum emission in the strongly lensed but otherwise typical SMG, SMM\,J21352$-$0102. There being no evidence of flux having been significantly resolved out on long baselines, we consider the new \co\ \jonezero\ flux density, luminosity and gas mass estimates -- within 5\% of the values measured by \citet{danielson11} based on single-dish GBT measurements -- to be minor revisions to those figures.

Taking full advantage of the $\sim 0.3''$ resolving power of the VLA's BnA configuration in $Ka$-band, we have identified four dense gas clumps in the \jonezero\ data cube, with counterparts mirrored about the lensing critical curve. The velocity offsets of the \co\ \jonezero\ line profiles of these clumps relative to the systematic velocity strongly suggest that they are related to the clumps $X$, $Y$, $Z_1$ and $Z_2$ identified by \citet{danielson11, danielson13}, whose analysis was based on the kinematic decomposition of the \co\ spectra in to multiple Gaussian components. 

We have investigated the \co\ excitation conditions within these $\sim100$\,pc sub-structures, determining the likely gas temperatures and molecular hydrogen volume densities through modelling of their \co\ SLEDs. This has allowed us to place plausible upper-limits on the value of the \co-to-H$_2$ conversion factor $\alpha_{\rm CO}$ for each of the clumps, which fall between those normally taken for ULIRGs and Milky Way GMCs. Depending on whether we adopt the firm upper or lower limits on the total molecular gas in SMM\,J21352, we find that these clumps may contain between 10--60\% of the total molecular gas in the galaxy.

\subsection{The FIRRC/spectral index relation as an age estimator}\label{sect:bressan}
Combining our updated SMA 870\,$\mu$m map with brand-new 4--8\,GHz $C$-band continuum observations of SMM\,J21352, we measure $q_{\rm IR}$, and hence test whether SMM\,J21352 obeys the far-infrared/radio correlation. We measure $q_{\rm IR}=2.39\pm0.17$ for SMM\,J21352, with individual pixels spanning a narrow range $q_{\rm IR}\sim 2.1$--$2.6$, which may have some interesting implications for the age distribution of material in the ISM. According to the \citet{bressan02} models of starburst galaxies, slight deviations from the FIRRC are expected for star-forming systems whose luminosity-weighted age is sufficiently young. Whereas the dust heating which produces thermal infrared emission is nearly instantaneous with the onset of star-formation, the radio spectrum is at early times comprised only of (relatively weak, flat-spectrum) thermal H{\sc ii} emission, and only at later times do the first supernovae -- which produce (steep spectrum) non-thermal synchrotron radio emission -- occur. The net effect of these differing time-scales is that both $\alpha_{\rm radio}$\ and $q_{\rm IR}$ for a typical galaxy may go through three key phases: (i) the \textit{proto-starburst phase}, during which dust-heating by young, massive stars rapidly elevates $q_{\rm IR}$, and the radio brightness is low, dominated by thermal emission; (ii) the \textit{evolving starburst phase}, during which time the heating of dust by low--intermediate mass stars continues to elevate $S_{\rm IR}$, but is out-paced by the much more rapid elevation in $L_{\rm radio}$ due to synchrotron emission originating in the first supernova remnants, and; (iii) the \textit{post-starburst/quiescent phase}, which is much longer-lived than either of the preceding phases, and occurs as the starburst event draws to a halt, and the heating of dust and the production of supernovae reach equilibrium, whereupon $q_{\rm IR}$ and $\alpha$ asymptote.

If the \citet{bressan02} models accurately describe the far-infrared and radio emission in SMM\,J21352, then it is notable that we see so few pixels with $q_{\rm IR}<2.2$, occupying the ``evolving starburst'' phase. The observed distributions in $\alpha_{\rm radio}$\ and $q_{\rm IR}$ appear to overlap the model tracks at both the young, proto-starburst ($<30$\,Myr) and the old, post-starburst ($>200$\,Myr) ends. While it is certainly possible that the ISM in SMM\,J21352 contains material of different ages, we consider the bi-modal age distribution implied by the scatter of the data about the model tracks to represent an implausible scenario. Given the representative error bars shown in Figure\,\ref{fig:alphavqir}, we cannot rule out the possibility that all the pixels have $\alpha_{\rm radio}$\ and $q_{\rm IR}$ values that lie at the young ends of the \citet{bressan02} evolutionary tracks; indeed, given the high instantaneous SFR in SMM\,J21352 \citep[$\sim 400$\,M$_{\odot}$\,yr$^{-1}$;][]{ivison10a}, we may expect to measure a low average age for material in the ISM. That the data are plausibly consistent with the \textit{entire} ISM having an age $<30$\,Myr \citep[there being no data covering the ``evolving starburst'' region of the ][model tracks with $q_{\rm IR}<2.2$]{bressan02} may indicate that SMM\,J21352 is a near-pristine, proto-starburst galaxy, whose high SFR has been rapidly triggered by a recent interaction.

An alternative explanation for the spread in $q_{\rm IR}$ seen within SMM\,J21352 is that it may be due to differing diffusion timescales and mechanisms for cosmic ray electrons and UV photons in the ISM. \citet{murphy06} find similar offsets in the logarithmic 24\,$\mu$m and 70\,$\mu$m-to-radio surface brightness ratios for $\sim$\,kpc regions within the discs of four nearby star-forming galaxies. These offsets disappear when the infrared maps are smoothed with exponential kernels, which attempt to replicate empirical ``leaky box'' cosmic ray electron models \citep[e.g.][]{bicay90, murphy06}. In these models, cosmic ray electrons escape from the young proto-stellar environment more quickly than do the thermal infrared photons, which can remain bound up in dust knots for $>100$\,Myr before eventually escaping via a random walk process. This leads to the radio emission having a correspondingly larger scale length ($\sim 1$--5\,kpc, \textit{cf.} $\sim 100$\,pc for the far-IR) and less clumpy structure than the infrared emission. Similar kpc-scale offsets in the FIRRC are seen in clumps in nearby spiral galaxies \citep{dumas11, basu12}.

There are thus two competing effects which may explain the pixel-to-pixel distribution in $q_{\rm IR}$ and $\alpha_{\rm radio}$ within SMM\,J21352, which operate on different time-scales and imply opposing estimates for the ages of the ISM: (i) that the offsets are due to the $\sim$10\,Myr lag between dust heating and the onset of the first supernovae, and reveal the existence of a young, near-pristine ISM; or (ii) that the offsets are due to the cosmic ray electron diffusion time-scale being significantly shorter than that of the infrared photons, which can remain bound up in dust knots for $>100$\,Myr before eventually escaping via a random walk process, implying an evolved ISM. The present data do not have the ability to fully distinguish between these two scenarios, however we can hope to revisit this dichotomy in future if we can obtain an estimate of the time-scale on which radiation pressure and mechanical feedback could conspire to disrupt the dense gas clumps in SMM\,J21352. Attempts to understand of the details of how galactic velocity fields are coupled to the turbulent dynamics of their constituent clouds -- a necessary ingredient of such an age estimate -- do, however, remain an area of ongoing research even in the local Universe \citep[][and references therein]{krumholz10}.

From these arguments, we see that the observed variations in $q_{\rm IR}$ for the star-forming clumps within SMM\,J21352 are not unexpected, however a more rigorous investigation of the spatially-resolved FIRRC will require forthcoming, high-resolution ALMA observations, which will allow us to create spatially-resolved far-infrared SEDs for each of the clumps -- and so obtain robust estimates of $L_{\rm IR}$ for regions within SMM\,J21352. This will eventually free us from the assumption that the SMM\,J21352 far-IR SED can simply be scaled to match the 870\,$\mu$m flux density of each clump.

\subsection{The star-formation law in $z\sim 2.3$ GMC complexes}
We have used the new \co\ \jonezero\ VLA map in conjunction with the SMA 870\,$\mu$m map and existing mid-$J$ \co\ maps from PdBI to investigate the properties of the cold molecular gas inside each clump, and ultimately place both the galaxy as a whole, and the clumps on the Schmidt-Kennicutt plot. The implied upper-limits on the \co-to-H$_2$ conversion factor, $\alpha_{\rm CO}$, derived via radiative transfer calculations are consistent with those previously derived for SMGs, though these limits are sensitive to the sizes estimated for the clumps. Within the limitations of these assumptions, we see that SMM\,J21352 and the clumps lie within the scatter of the $L_{\rm IR}/L'_{\rm CO}$ plot, at the moderate and low luminosity ranges, respectively, in good agreement with the findings and implications of recent \co\ \jonezero-based surveys \citep[e.g.][]{harris10, ivison11, riechers11b, thomson12}. While these clumps appear to be marginally offset from the ``normal'' star formation law claimed for $BzK$ galaxies on the $\Sigma_{\rm SFR}/\Sigma_{\rm gas}$ surface density plot, we note that: (i) $M_{\rm gas}$ is usually very poorly constrained in $BzK$ galaxies due to the reliance on an assumed brightness temperature ratio to predict the \co\ \jonezero\ line flux from mid-$J$ observations, and on the subsequent application of a standard $\alpha_{\rm CO}$ to all $BzK$ galaxies \citep{daddi10a}, leaving scope for considerable variation in the values of $\Sigma_{\rm SFR}$ and $\Sigma_{\rm gas}$ in individual $BzK$ galaxies that is not reflected in the scatter in Figure\,\ref{fig:sk-eyelash}; (ii) using a lower $\alpha_{\rm CO}$ for SMM\,J21352 -- which is allowed by our radiative transfer modelling -- would imply a higher gas mass surface density (at a fixed $\Sigma_{\rm SFR}$), and so would shift the clumps to the right in Figure\,\ref{fig:sk-eyelash} and in to better agreement with a single star formation law.

\subsection{The relationship between \co\ excitation and star formation}
Finally, we have used our suite of multi-wavelength data to test the prediction of a recent theoretical model \citep{narayanan14} which claims that although the \co\ excitation conditions in galaxies are determined by a morass of difficult-to-observe quantities (including temperature, density and optical depth), each of these correlates in some predictable way with the SFR surface density, $\Sigma_{\rm SFR}$, such that a measurement of $\Sigma_{\rm SFR}$ can yield a prediction for the shape of the \co\ SLED. Using SMA 870$_{\mu {\rm m}}$ observations of SMM\,J21352, in conjunction with the well-sampled far-IR SED of the galaxy, we measure $L_{\rm IR}$, SFR and $\Sigma_{\rm SFR}$ for each of the dense molecular gas clumps identified in \co\ \jonezero, and create a prediction for the \co\ SLED. We find that in three of the four clumps, the predicted SLEDs are in good agreement with those derived from multi-$J$ \co\ imaging, providing tentative support for the \citet{narayanan14} model of \co\ excitation in star-forming galaxies.

\section*{Acknowledgements}
APT and IRS acknowledge support from STFC (ST/I001573/1). AMS gratefully acknowledges an STFC Advanced Fellowship through grant number ST/H005234/1. IRS also acknowledges a Leverhulme Fellowship, the ERC Advanced Investigator Programme {\sc dustygal} (\#321334) and a Royal Society/Wolfson merit award. RJI acknowledges support from the ERC in the form of Advanced Grant {\sc cosmicism}. The National Radio Astronomy Observatory is a facility of the National Science Foundation operated under cooperative agreement by Associated Universities, Inc. Based on observations carried out with the IRAM Plateau de Bure Interferometer. IRAM is supported by INSU/CNRS (France), MPG (Germany) and IGN (Spain). The Submillimeter Array is a joint project between the Smithsonian Astrophysical Observatory and the Academia Sinica Institute of Astronomy and Astrophysics and is funded by the Smithsonian Institution and the Academia Sinica. The authors thank the anonymous referee for their considered and helpful comments and suggestions, which have improved the quality of this paper.

\FloatBarrier
\bibliographystyle{mnras} 
\bibliography{references}
\bsp

\label{lastpage}

\end{document}